\documentclass[twocolumn,a4]{revtex4-1}
\usepackage{relsize}
\usepackage[countmax]{subfloat}
\usepackage{graphicx}
\usepackage{amssymb, amsmath,amssymb,amsfonts}
\usepackage{amsthm,mathrsfs,amsopn}
\usepackage{dcolumn}
\usepackage{bm}
\usepackage{color}
\usepackage[utf8]{inputenc}
\usepackage[table]{xcolor} 
\usepackage{floatrow}
\usepackage{lettrine}

\begin{document}

\title{A universal route to {pattern formation}}

\author{Malbor Asllani$^{1}$, Timoteo Carletti$^{2}$, Duccio Fanelli$^{3}$, Philip K. Maini$^{4}$ \vspace*{.25cm}}
\affiliation{$^1$MACSI, Department of Mathematics and Statistics, University of Limerick, Limerick V94 T9PX, Ireland}
\affiliation{$^2$Department of Mathematics and naXys, Namur Institute for Complex Systems, University of Namur,
rempart de la Vierge 8, B 5000 Namur, Belgium}
\affiliation{$^3$Dipartimento di Fisica e Astronomia, Universit\`{a}  di Firenze, INFN and CSDC, Via Sansone 1, 50019 Sesto Fiorentino, Firenze, Italy}
\affiliation{$^4$Mathematical Institute, University of Oxford, Woodstock Rd, OX2 6GG Oxford, UK}

%

\maketitle

\noindent
\textbf{Self-organization, the ability of a system of microscopically interacting entities to shape macroscopically ordered structures, is ubiquitous in Nature. Spatio-temporal patterns are abundantly observed in a large plethora of applications, encompassing different fields and scales. Examples of emerging patterns are the spots and stripes on the coat or skin of animals~\cite{angelfish,leopard}, the spatial distribution of vegetation in arid areas~\cite{vegetation}, the organization of the colonies of insects in host-parasitoid system{s}~\cite{insectpattern} and the architecture of large complex ecosystems~\cite{ecopattern}. Spatial self-organization can be described following the visionary intuition of Alan Turing, who showed how {non}-linear interactions between slow diffusing activators and fast diffusing inhibitors could induce patterns~\cite{prigogine,murray}. The Turing instability, as the mechanism {described} is universally referred to, was raised to paradigm status in those realms of investigations where microscopic entities are subject to diffusion, from small biological systems to large ecosystems.  Requiring a significant ratio of the assigned diffusion constants  however {is} a stringent constraint, which limited the 
applicability of the theory. Building on the observation that {spatial} interactions are usually direction biased, and  often {strongly asymmetric}~\cite{asym_geometric,osmosis,murray,dallon,universal}, we here propose a novel framework  
for the generation of short wavelength patterns which overcomes the limitation inherent in the Turing formulation. In particular, we will prove that patterns can always set in when the system is composed by sufficiently many cells {--} the units of spatial patchiness {--} and for virtually any ratio of the diffusivities involved. Macroscopic patterns that follow the onset of the instability are robust and show oscillatory or steady state behaviour.}

\begin{figure*}[t]
\centering
\includegraphics[width=0.9\textwidth]{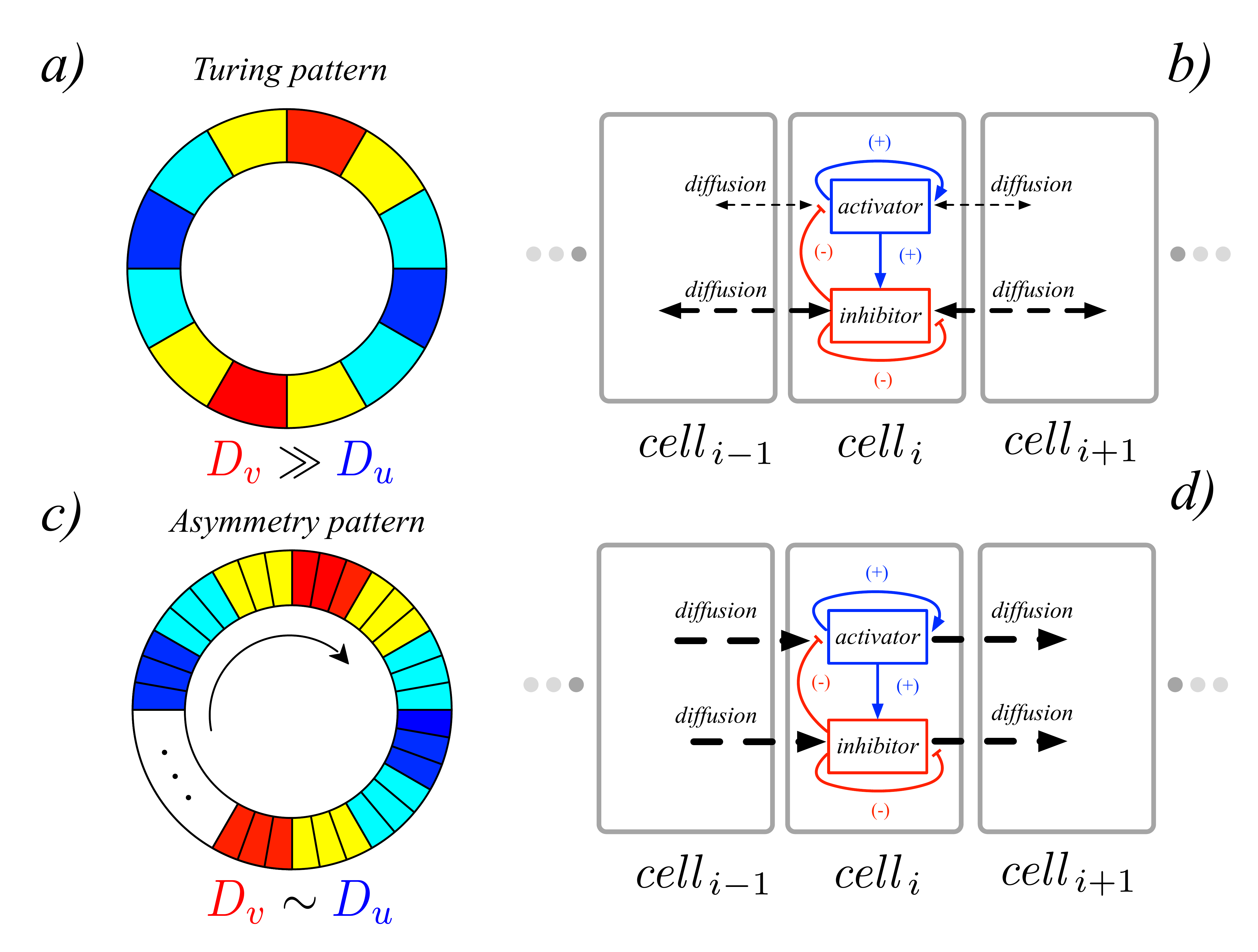}
\caption{\textbf{Conventional Turing instability vs. the asymmetry-driven model: a schematic representation.} $\textbf{(a)}$ In Turing's original model the onset of pattern formation is studied for a two species model reacting and diffusing on a collection of cells, arranged so as to form a $1D$ ring. $\textbf{(b)}$ Turing instability requires breaking the symmetry among the species. In particular, feedback loops (positive for the activators and negative for the inhibitors) are {proposed}. Further,  the inhibitor should relocate in space faster than the activator, $D_v\gg D_u$. The diffusion between neighbor{ing} cells is assumed symmetric. $\textbf{(c)}$ In the asymmetry-induced instability instead the system is made up of a larger number of cells and the diffusion is asymmetric, as schematised by the counterclockwise arrow. $\textbf{(d)}$ The instability is triggered by increasing the number of cells, for virtually any ratio of the diffusion constants, provided the latter take sufficiently large values.}
\label{fig:patterns}
\end{figure*}

\noindent
In the early $1950's$, Alan Turing  laid down, in a seminal paper \cite{turing}, the mathematical basis of pattern formation, the discipline that aims at explaining the richness and diversity of forms displayed in Nature. Turing{'s} idea paved the way {for} a whole field of investigation and fertilized a cross{-}disciplinary perspective to yield a universally accepted paradigm of self-organization~\cite{ball}. The onset of pattern formation on a bound spatial domain originates from the loss of stability of an unpatterned equilibrium. To start with, Turing proposed a minimal model composed of at least two chemicals, {hereby} termed species. {The} species were assumed to diffuse across an ensemble of cells, adjacent to each other and organized in {a} closed ring, as depicted in Figure~\ref{fig:patterns} a).  One of the species should trigger its own growth, acting therefore as a self-catalyst. This is opposed by the competing species,  which therefore promotes an effective stabilization of the underlying dynamics. The emergence of the ensuing spatial order relies on these contrasting interactions and requires, as {an} unavoidable constraint, a marked difference (as measured by the ratio) of the diffusion constants associated {with} the interacting species~\cite{murray}. This symmetry-breaking mechanism is at the core of a general principle, widely known as the Turing instability~\cite{prigogine}: small inhomogeneous perturbations from a uniform steady state initiate the instability and individuals, in a quest for space and resources, organise in spatially extended, regular motifs~\footnote{{The loss of stability of an homogeneous equilibrium, as triggered by an external perturbation, will be referred to as a Turing instability, as an natural extension of the  original model discussed in~\cite{turing}}}. This original idea was built upon by Meinhardt~\cite{Meinhardt}, who proposed the notion of activators and inhibitors, so that the Turing patterning principle could be conceptualized as {arising through} short-range activation (slow diffusion), long-range inhibition (fast diffusion).

Pattern formation for systems evolving on cellular arrays was further analyzed by Othmer and Scriven~\cite{othmer} under the assumption of symmetric diffusion. {Agregates} of cells yield macroscopic tissues which, in general, can be schematized, to a reasonable approximation, by regular lattices~\cite{murray}. Branching architectures, or coarse{-}grained models of compartimentalized units, justify invoking the generalized notion of {a} spatial network. In this case, the nodes stand for individual cellular units, linked via a heterogen{e}ous web of intertangled connections, as exemplified by the network structure. The study of pattern formation for reaction-diffusion systems anchored on symmetric networks was develop{ed} in~\cite{nakao}. Diffusion instigates a spatial segregation of species, a counterintuitive outcome of Turing analysis, which holds true both in its continuous and discrete (lattice or network based) versions. 
 
Notwithstanding its unquestionable conceptual relevance, it is still unclear if the Turing scheme, in its original formulation, can be successfully applied to describe real life systems. The large ratio of (inhibitor vs. activator) diffusivities can be attained in {the} laboratory, under controlled operating conditions~\cite{dekepper}. An unequivocal proof of {the} existence of Turing patterns in non{-}artificial, biological or ecological, contexts is however still lacking, despite several candidate systems {that} have been thoroughly scrutinized and challenged in this respect~\cite{JMB,membrane}.  

Asymmetries in flows{,} as displayed in real systems {via} chemical or electrical gradients~\cite{asym_geometric}, can possibly provide the key to bridge the gap between theory and applications. {For example, o}smosis~\cite{asym_geometric,osmosis} in the cell membrane or chemotaxis in the motility of cells~\cite{murray,dallon} are examples of asymmetric transport. The case of {\emph{Dictyostelium}} is also {worth} mentioning: this is a multi-celled eukaryotic bacterivore, which develop{s} pseudopodia under externally induced chemotaxis, so triggering a directed biased motion~\cite{dyctostelium,dallon}. Rovinsky \& Menzinger~\cite{menzinger_theory,menzinger_exp} proved that {an unbalance in the directions of the flows of the species} can destabilise the spatially homogeneous state of the system. The differential-flow-induced instability yields oscillatory patterns for a sufficiently pronounced degree of asymmetry. This mechanism suffers{,} however{,} {from} the same limitation that applies to the standard Turing theory: diffusion constants need to be significantly different for the pattern to emerge. It is however crucial to realize that asymmetric diffusion can eventually yield a richer pattern formation dynamics, as intuited by Othmer and Scriven  in their pioneering work~\cite{othmer}. Turing theory for systems evolving on directed networks was recently cast on rigorous grounds in ~\cite{asllani} and asymmetry in the diffusion shown to produce a consistent enlargement of the region of Turing-like patterning.

Starting from these premises, we {prove} that patterns can develop for virtually any ratio of the diffusivities, larger or smaller than unit{y}, assuming asymmetric transport on (i) an array of sufficiently many cells and (ii) with adequately large values of the nominal diffusion constants (while keeping the sought ratio fixed). To anticipate some of the technical aspects to be addressed in the following, we will show that tuning the diffusivity of just one species -- the ratio of the two diffusivities being frozen to a constant strictly different from (but arbitrarily close to) one -- amounts to performing a homothetic transformation of the spectrum of the generalized Laplacian operator in the complex plane. The Laplacian eigenvalues consequently move along straight lines, the slope being set by the number, $\Omega$, of cells that define the lattice space. By modulating these latter quantities, one can {\it always}  get the eigenvalues to protrude into the region of instability. The patterns that follow the onset of the instability are either oscillatory or stationary. A pictorial representation of the proposed scheme is provided in panels c) and d) {of} Figure~\ref{fig:patterns}. Panels a) and b) refer instead to the conventional scenario which assumes undirected transport.

Let us begin by considering a generic two species model of the reaction-diffusion type. The concentration of the two species are{,} respectively{,} labelled $u_i$ and $v_i$, where the index $i$ refers to the hosting cell and $i$ runs from $1$ to $\Omega$.  The governing equations can be cast in the form:
\begin{eqnarray}
\frac{du_i}{dt} &=& f(u_i,v_i) +  D_u \left[ \sum_{j=1}^\Omega A_{ij} u_j - \sum_{j=1}^\Omega A_{ji} u_i \right] \nonumber\\
\frac{dv_i}{dt} &=& g(u_i,v_i)+ D_v \left[ \sum_{j=1}^\Omega A_{ij} v_j - \sum_{j=1}^\Omega A_{ji} v_i \right]
\label{eq:RD}
\end{eqnarray}
where $D_u$ and $D_v$ stand for the diffusion constants and $f(\cdot, \cdot)$, $g(\cdot, \cdot)$ are the nonlinear reaction terms that {model} the local (on site) dynamics of the species. $A_{ij}$ are the binary entries of the adjacency matrix that specifies the topology of the embedding spatial support. For the case at hand (panel c) of Fig. \ref{fig:patterns}), the adjacency matrix $\boldmath{A}=\{ A_{ij} \}$ is circulant (i.e. invariant for translation). The transport operator {introduced} here represents the straightforward generalization of standard diffusion, to the case where the spatial arrangements of the cells is supposed heterogeneous, either in terms of physical links, connecting individual patches of the collection, or in terms of their associated weights. Both are viable strategies for imposing the sought asymmetry which sits at the core of the mechanism upon which we shall hereby elaborate. In short, the hypothesized spatial coupling implements a local balance of incoming and outgoing fluxes, as seen from the observation cell $i$. {We} introduce the Laplacian operator ${\mathbf{L}}$ with entries $L_{ij}=A_{ij}-k_i^{out}\delta_{ij}$ where $\delta_{ij}$ denotes the Kronecker function and $k_i^{out}=\sum_{j=1}^{\Omega} A_{ji}$ quantifies the outgoing degree of cell $i$. The contributions that relate to inter-cell couplings in Eqs.  (\ref{eq:RD}) can be{,} respectively{,} rewritten in the equivalent, more compact form $\sum_{j=1}^\Omega D_u {L}_{ij} u_j$ and $\sigma \sum_{j=1}^\Omega D_u {L}_{ij} v_j$, where $\sigma$ represents the ratio of the diffusion constants $D_u$ and $D_v$. In the following, we will denote with the symbol  $\boldsymbol{\mathcal{L}} $ the Laplacian $\mathbf{L}$, modulated by the multiplicative constant $D_u$, namely $\boldsymbol{\mathcal{L}}  \equiv D_u \mathbf{L}$. As a key observation for what follows, we emphasize {that} the magnitude of {the} eigenvalues of $\boldsymbol{\mathcal{L}}$ can be freely controlled by {the value of} $D_u$. Changing $D_u$ (while for instance keeping $\sigma$ frozen) implies performing a homothetic transformation of the spectrum of $\mathbf{L}$, as we shall hereafter clarify.

\begin{figure*}[t]
\centering
\includegraphics[width=0.9\textwidth]{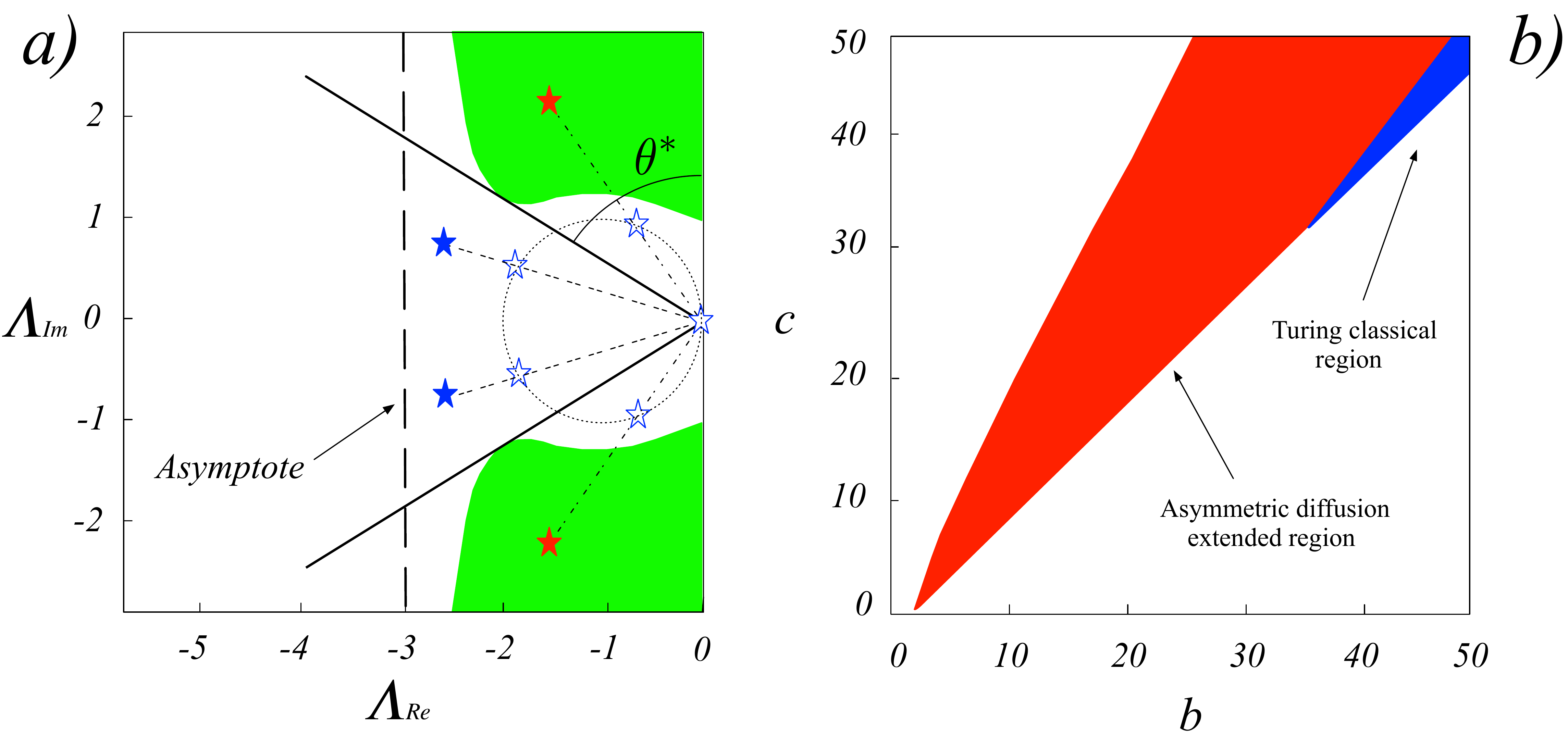}
\vspace*{-.25cm}
\caption{\textbf{On the mechanism of pattern formation with asymmetric diffusion.} $\textbf{a)}$ The region colored in green {denotes} the domain of the complex plane $\textbf{z}=\big(\Lambda_{Re},\Lambda_{Im}\big)$ where the instability can eventually take place. This is a pictorial representation of a general situation that is always found, irrespectively of the specific choice of the reaction model. The blue empty symbols stand for the spectrum of the Laplacian operator $\mathbf{L}$ obtained from a directed lattice of the type depicted in Fig. \ref{fig:patterns} c). The eigenvalues are distributed on the unitary circle, as discussed in the Method{s} section. When increasing  $\Omega$, the number of cells that compose the examined lattice, the Laplacian spectrum displays eigenvalues with progressively larger (but still negative) component $\Lambda_{Re}$ and still lying on the circle. We {denote by} $\theta$ the inclination of the line ({dot} dashed in the figure) that connects the selected eigenvalue (empty cross) to the origin of the complex plane. This latter eigenvalue can be freely moved along the aforementioned line, by modulating the diffusion coefficient $D_u$ while keeping the ratio $\sigma$ fixed. Stated differently,  the spectrum of the matrix $\boldsymbol{\mathcal{L}}$ is an homothetic transformation (with centre $0$ and ratio $D_u$) of the spectrum of $\mathbf{L}$. If $\theta$ is sufficiently small, i.e. if the dashed line is steeper than the solid one (the instability threshold), it is always possible (by increasing $D_u$) to move the associated eigenvalue inside the region (green) {of} the instability (see the filled red star). If the inclination $\theta$ is larger than the critical one, $\theta^*$, the eigenvalue{s} that slide on the corresponding dashed line are permanently confined outside the domain of instability (filled blue star). The vertical dashed line is an asymptote and sets the leftmost boundary of the instability domain, as described in the annexed Methods.  $\textbf{b)}$ To provide a quantitative illustration of the method, we consider the Brusselator model, $f(u,v)=1-(b+1)u+c u^2 v$ and $g(u,v)=bu-cu^2v$, where $b$ and $c$ are free control parameters. In the plane $(b,c)$, we isolate the domain where the conventional Turing instability takes place (small blue domain), for $\sigma=1.4$.  The region shaded in red identifies the domain of instability that is found  when the system is evolved on a directed lattice made of  $\Omega=1000$ cells and assuming $D_u=100$. The ratio $\sigma$ is kept constant to the reference value $1.4$.}
 \label{fig:new_theory}
\end{figure*}

Assume now that the reaction dynamics admits a stable fixed point $({u}^*, {v}^*)$, namely that $f({u}^*, {v}^*)=g({u}^*, {v}^*)=0$. Hence, the spatially extended system (\ref{eq:RD}) possesses a homogeneous equilibrium solution $(\textbf{u}^*, \textbf{v}^*)$ which is obtained by replicating on each of the $\Omega$ cells the solution $({u}^*, {v}^*)$. This conclusion can be readily derived by noticing that, by definition,  $\sigma \sum_{j=1}^\Omega \mathcal{L}_{ij} K =0 $, for any constant $K$ (recall that the system is hosted on a lattice with periodic boundary conditions).  The homogeneous fixed point can{,} in principle{, become} unstable upon injection of a tiny heterogeneous perturbation, as in the spirit of the original Turing mechanism. The conditions for the emergence of the instability are determined by a linear stability analysis that follows the procedure outlined in ~\cite{asllani} and that we develop in the annexed Methods. The calculation yields the compact inequality (\ref{eq:instdom}) which can be graphically illustrated in the complex plane $\textbf{z}=\big(\Lambda_{Re}, \Lambda_{Im}\big)$, where the eigenvalues $\Lambda^{(\alpha)}$ of the Laplacian operator $\mathbf{L}$ reside (see Methods). In fact, inequality  (\ref{eq:instdom}) enables one to delimit a model-dependent region of instability, which is depicted in  panel a) of Fig.~\ref{fig:new_theory}. When drawing the domain of interest, we assumed an abstract setting, without insisting on the specific details that stem from a particular reaction model. For what will follow, it is only important to appreciate that condition (\ref{eq:instdom})  makes it possible to define a portion of the parameter space, by construction symmetric with respect to the horizontal (real) axis, that is eventually associated {with} the onset of the instability. More specifically, if a subset of the spectrum of the Laplacian $\boldsymbol{\mathcal{L}}$ fall{s} inside the region outlined above, the instability can take place (red stars in panel a) of Fig.~\ref{fig:new_theory}).  This conclusion is  general and {independent} of the reaction scheme employed. Notice that when the two regions merge and incorporate a finite part of the real 
axis, the outbreak of the instability is also possible on a symmetric support (when i.e. the eigenvalues of the Laplacian operator are real).

\begin{figure*}[t]
\includegraphics[width=1\textwidth]{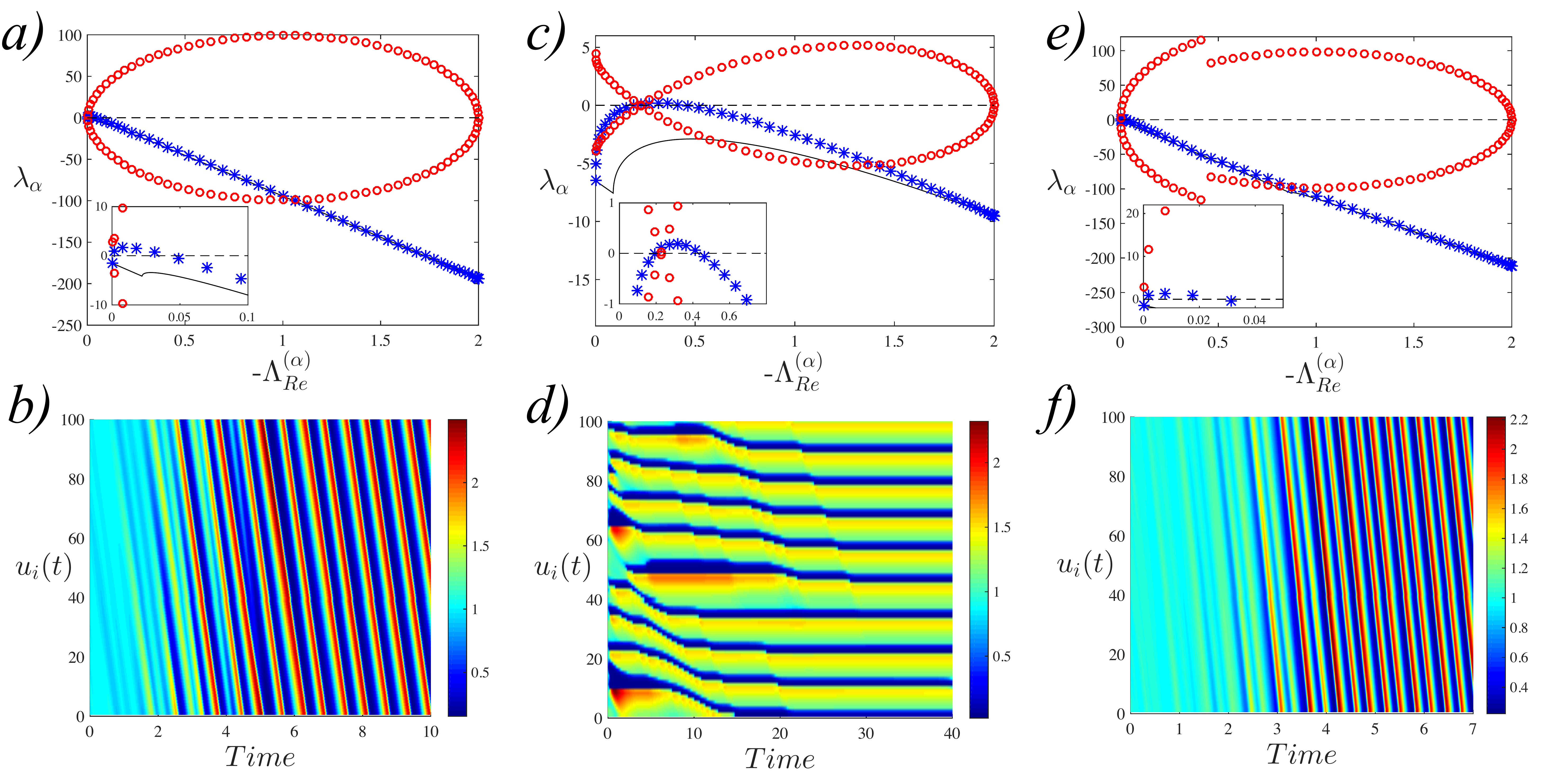}
\caption{\textbf{Patterns triggered in the Brusselator model by asymmetric diffusion on a $1D$ ring.} Left column: real (blue stars) and imaginary (red circles) dispersion relation $\lambda_{\alpha}$ $\textbf{a)}$ and the associated pattern evolution $\textbf{b)}$. The ensuing pattern is oscillatory and organizes as a travelling wave. The parameters are set to $b=8$, $c=10$, $D_u=100$, and $\sigma=1.4$. Center column: dispersion relation $\textbf{c)}$ and pattern evolution $\textbf{d)}$ yielding a {steady state pattern}, when $b=50$, $c=62$, $D_u=10$ and  $\sigma=1.4$.  Right column: dispersion relation $\textbf{e)}$ and pattern evolution $\textbf{f)}$ yielding a traveling wave with $b=8$, $c=10$, $D_u=140$ and $\sigma=1/1.4 \simeq 0.71$. In the insets in the upper panels, we zoom {on} the (real and imaginary) dispersion relations focusing on the portion of the curve where the instability takes place. In all cases, $\Omega=100$. Notice that, for the Brusselator model,  oscillatory and stationary patterns are numerically detected for $\sigma>1$, while for $\sigma<1$ oscillatory patterns are solely found. There is{,} however{,} no reason to exclude that other reaction schemes would {yield} stationary stable patterns for $\sigma<1$ as an outcome of the scheme {shown} here.}
\label{fig:patt}
\end{figure*}

We are now in a position to elaborate on the universal mechanism which drives a reaction system unstable, when bound to  performing asymmetric diffusion on a discrete collection of lattice sites. As we shall realize, considering a finite, although large, ensemble of mutually connected cells is one of the key ingredients that instigates the instability {in an activator-inhibitor system for virtually any reaction parameters and} any ratio of the diffusi{on coefficients} {(including $\sigma < 1$ and except for the zero measure, limiting condition $\sigma=1$}).  We begin by observing that the Laplacian matrix $\mathbf{L}$, associated {with} a closed directed ring as assumed in Fig. \ref{fig:patterns} c), is {a} circulant matrix. This simplified geometrical arrangement is solely assumed for illustrative purposes. Our conclusions hold in general and include those systems that undergo asymmetric diffusion on lattices in arbitrary dimensions, subject to periodic boundary conditions. The eigenvalues of the Laplacian $\mathbf{L}$  (see Methods) are complex and fall on the unitary circle centered at $(-1,0)$ {--} empty stars in Fig. \ref{fig:new_theory}, panel a). When making $\Omega$ larger, one progressively reduces the spectral gap, the relative distance between the two Laplacian eigenvalues that display the largest real parts. Recalling that the largest eigenvalue of the Laplacian is by definition zero, this implies that the second eigenvalue of  $\mathbf{L}$ (ranked in descending order, with respect to the value of their associated real parts) tends to approach the origin of the complex plane, when $\Omega$ {is} increased \footnote{The eigenvalues comes in conjugate pairs. In what it follows we shall refer to the eigenvalue in the pair that displays positive imaginary part.}. This{,} in turn{,} implies that we can control at will the inclination (a measure complementary to the slope) $\theta$ of the line (dashed in the figure) that connects the second largest eigenvalues to the origin of the complex plane. Similar considerations apply to the other eigenvalues that follow in the ranking. {In  Fig. \ref{fig:new_theory}) a) we label by $\theta^*$ the critical inclination of the solid line that intersects tangentially the domain of instability traced according to inequality (\ref{eq:instdom})}.  Clearly, by construction, a symmetric line always exists with inclination $-\theta^*$, that is tangent to the region of instability in the lower complex semi-plane. Assume now that $\Omega$ is sufficiently large so that $\theta < \theta^*$. Then recall that for the instability to emerge, at least one eigenvalue of the rescaled Laplacian $\boldsymbol{\mathcal{L}}$ has to protrude into the region where the instability is bound to occur. On the other hand, the eigenvalues of $\boldsymbol{\mathcal{L}}$ are a homothetic transformation, with centre $0$ and ratio $D_u$, of the eigenvalues $\mathbf{L}$. In other words, we can force the second eigenvalue of $\mathbf{L}$ to move along the line {to which} it belongs by modulating the diffusion constant $D_u$, while keeping $\sigma$ constant, and so invade the region {of} the instability (filled red star). If  $\theta > \theta^*$, {then} the eigenvalue is instead confined {to} the stable portion of the complex plane for each choice of the scaling factor $D_u$. {In summary,} the system can be triggered unstable by simultaneously acting on two independents {``}knobs{''}: first, by making $\Omega$ sufficiently large, we force a subset of eigenvalues {into} the vicinity of the origin, thus making their associated $\theta$ smaller that the critical amount $\theta^*$. Secondly, by acting on $D_u$ (and consequently on $D_v$ so as to maintain $\sigma$ unchanged), we make the eigenvalues slide on their corresponding lines, until the instability threshold is eventually breached. Interestingly, the instability involves long wavelengths (hence yielding macroscopic patterns) since, by construction, the real part of the eigenvalues associated {with} the modes triggered {to grow} approaches zero, when $\Omega$ increases.

In Fig. \ref{fig:new_theory} b) we present results for a specific case study, the Brusselator model, often invoked in the literature as {a paradigm} nonlinear reaction scheme for studying  self-organised phenomena, synchronisation, Turing patterns and oscillation death. {We find the region in the parameter space} $(b,c)$ {for which the homogeneous steady state is} unstable for $\sigma=1.4$. The Brusselator may undergo a conventional Turing instability, in the portion of the plane that is colored in blue. The asymmetry driven instability (for $\Omega=1000$ cells and assuming $D_u=100$) is found on a considerably larger domain, which can be made arbitrarily {large} by further modulating $\Omega$ and $D_u$. 

In Fig.~\ref{fig:patt}, panels b), d) and f), we {show} patterns found when integrating the Brusselator model for different parameter {values}. Panels a), c) and e) display the real and imaginary parts of the eigenvalues $\lambda_{\alpha}$ (the dispersion relation, see Methods) against {$-\Lambda^{(\alpha)}_{Re}$}. When the imaginary component of $\lambda_{\alpha}$ is small as compared to the corresponding real part, {steady state} patterns are found to emerge. Otherwise, the patterns are oscillatory in nature. This is a rule of the thumb, which seems to implement a necessary criterion. The mechanism of pattern selection is in fact heavily influenced by {the} nonlinearities, which become prominent beyond the initial stages of evolution, hence limiting the predictive ability of the linear stability analysis. It is worth stressing that the patterns displayed in Fig.  \ref{fig:patt} occur for a choice of parameters for which the classical Turing instability is not permitted and for both $\sigma$ larger and smaller than one. In particular, one can analytically show that the asymmetry driven instability, that we here introduced and characterized, can develop for any $\sigma=1 \pm \epsilon$ with $0< \epsilon \ll1$. This conclusion holds in general, as can be proved rigorous{ly (see Methods)}. 

To sum{marise}, we have developed a general theory of pattern formation for the {case of asymmetric} transport of constituents in a large assembly of adjacent cells. Macroscopic oscillatory or steady state patterns {in an activator-inhibitor system} are {now} found for virtually any ratio of diffusivities, larger or smaller than {one,} {and any choice of the reaction parameters. The applications of this newly proposed route to pattern formation are multiple and include all those settings, from biology to ecology passing through neuroscience \cite{asym_geometric,asym_eco,osmosis,sporns,dallon,dyctostelium}, where asymmetric flows are reported to occur.}

\bibliography{Non-normal.bib}
\vspace*{-.5cm}
\section*{Acknowledgments}\vspace*{-.5cm}
{\small The work of M.A. was supported by a FRS-FNRS Postdoctoral Fellowship.}\\
\clearpage
%
%
%
%
%
\noindent{\textbf{{\Large Methods}}}\\
 \vspace{0.5truecm}

\noindent{\textbf{{The dispersion relation and the conditions for instability}}}\\
Lineari{sing} the dynamics of system (\ref{eq:RD}) around the fixed point solution $(\textbf{u}^*, \textbf{v}^*)$ {we obtain}:

\begin{eqnarray*}
\frac{d}{dt}\begin{pmatrix} \delta \textbf{u} \\ \delta \textbf{v} \end{pmatrix}&=&\begin{pmatrix}
f_u \boldsymbol{\mathcal{I}} + D_u \mathbf{L} & f_v \boldsymbol{\mathcal{I}}\\g_u \boldsymbol{\mathcal{I}} & g_v + D_v \mathbf{L} \end{pmatrix}\begin{pmatrix} \delta \textbf{u} \\ \delta \textbf{v}\end{pmatrix}\\&=& \big(\boldsymbol{\mathcal{J}} + \boldsymbol{\mathcal{D}}\big)\begin{pmatrix} \delta \textbf{u} \\ \delta \textbf{v}\end{pmatrix}
\end{eqnarray*}
where $\delta \textbf{u}$, $\delta \textbf{v}$ stands for the perturbation vectors. $\boldsymbol{\mathcal{J}}=\begin{pmatrix}
f_u \boldsymbol{\mathcal{I}} & f_v \boldsymbol{\mathcal{I}}\\g_u \boldsymbol{\mathcal{I}} & g_v \boldsymbol{\mathcal{I}} \end{pmatrix}$ is the Jacobian matrix, evaluated at the equilibrium point, stemming from the reaction terms and $\boldsymbol{\mathcal{D}}=\begin{pmatrix}
D_u  \mathbf{L} & \boldsymbol{\mathcal{O}}\\\boldsymbol{\mathcal{O}} & D_v \mathbf{L} \end{pmatrix}$, where $\boldsymbol{\mathcal{O}}$ is the $\Omega\times \Omega$ zero-valued matrix. Introduce the basis formed by the eigenvectors $\Phi_i^{(\alpha)}$, with $\alpha=1,\dots,\Omega$, of the Laplacian operator $ \mathbf{L}$. {We} have $\sum_j L_{ij} \Phi_j^{(\alpha)} = \Lambda^{(\alpha)} \Phi_i^{(\alpha)}$ where $\Lambda^{(\alpha)}$ identif{ies} the eigenvalues of $ \mathbf{L}$. Th{e} latter {operator} is asymmetric, as is {the} matrix $\mathbf{A}$, and thus $\Lambda^{(\alpha)}$ are, in principle, complex. Further,  the eigenvectors form an orthonormal basis, for the case at hand. Hence, to solve the above linear system, we expand the perturbation in terms of the basis of the eigenvectors, i.e. $\delta u_i= \sum_{\alpha=1}^\Omega b_{\alpha} \Phi_i^{(\alpha)}$ and $\delta v_i= \sum_{\alpha=1}^\Omega  c_\alpha \Phi_i^{(\alpha)}$. At this point it is straightforward to show that the $2\Omega \times 2\Omega$ system reduces to a $2\times 2$ eigenvalue problem, for each choice of the scalar index $\alpha=1,\dots,\Omega${:} 
\begin{eqnarray*}
\det\begin{pmatrix} \textbf{J}_\alpha - \lambda_\alpha\textbf{I}\end{pmatrix}&=&\det\begin{pmatrix}
f_u + D_u\Lambda^{(\alpha)}-\lambda_\alpha \hspace*{-.5cm}& f_v \\g_u  & g_v + D_v\Lambda^{(\alpha)}-\lambda_\alpha \end{pmatrix}\\&=&0
\end{eqnarray*}
where $\textbf{J}_\alpha$ is the $2\times 2$ modified Jacobian, {i.e.} $\textbf{J}_\alpha\equiv\textbf{J} + \textbf{D}\boldsymbol{\Lambda}^{(\alpha)}$ with $\textbf{D}=diag(D_u,D_v)$, $\boldsymbol{\Lambda}^{(\alpha)}=diag(\Lambda^{(\alpha)},\Lambda^{(\alpha)})$. The {steady state is} unstable to {small} heterogeneous perturbation{s}, if $\lambda_\alpha$ {has} a positive real part over a finite range of modes. The dispersion relation (the largest real part of $\lambda_\alpha$, for any given $\alpha$) can be readily computed from:
\begin{equation}
\lambda_\alpha=\frac{1}{2}\left[\left(\mathrm{tr}\textbf{J}_\alpha\right)_{Re}+\gamma\right] + \frac{1}{2}\left[\left(\mathrm{tr}\textbf{J}_\alpha\right)_{Im}+\delta\right]\iota
\label{eq:disp_rel}
\end{equation}
where 
\begin{eqnarray*}
\gamma&=&\sqrt{\frac{A+\sqrt{A^2+B^2}}{2}}\\
\delta&=&\mathrm{sgn}(B)\sqrt{\frac{-A+\sqrt{A^2+B^2}}{2}}
\end{eqnarray*}
and
\begin{eqnarray*}
A&=&\left[\left(\mathrm{tr}\textbf{J}_\alpha\right)_{Re}\right]^2 - \left[\left(\mathrm{tr}\textbf{J}_\alpha\right)_{Im}\right]^2-\left[\left(\det\textbf{J}_\alpha\right)_{Re}\right]^2\\
B&=&2\left(\mathrm{tr}\textbf{J}_\alpha\right)_{Re}\left(\mathrm{tr}\textbf{J}_\alpha\right)_{Im}-\left[\left(\det\textbf{J}_\alpha\right)_{Im}\right]^2.
\end{eqnarray*}
Here $\mathrm{sgn}(\cdot)$ is the sign function and $f_{Re}$, $f_{Im}$ indicate, respectively{,} the real and imaginary parts of the operator $f$. 
To further manipulate Eq.~(\ref{eq:disp_rel}), we make use of the definition of a square root of a complex number. Take, $z=a+b\iota$ {where $\iota=\sqrt{-1}$ is the imaginary unit}, then
\begin{equation*}
\sqrt{z}=\pm \left(\sqrt{\frac{a+|z|}{2}}+ \mathrm{sgn}(b)\sqrt{\frac{-a+|z|}{2}}\iota\right).
\end{equation*}

The instability sets in when 
$|\left(\mathrm{tr}\textbf{J}_\alpha\right)_{Re}|\leq \gamma$, a condition that translates into the following inequality \cite{asllani}: 
\begin{equation}
S_2(\Lambda_{Re}^{(\alpha)})[\Lambda_{Im}^{(\alpha)}]^2\leq -S_1(\Lambda_{Re}^{(\alpha)}),
\label{eq:instdom}
\end{equation}
where $(\Lambda_{Re}, \Lambda_{Im})$ span the complex plane where the Laplacian eigenvalues reside. In the above relation
$S_1$, $S_2$ are polynomials which take the following explicit form:
\begin{eqnarray*}
S_1(x)&=& C_{14}x^4+C_{13}x^3 + C_{12}x^2 + C_{11}x+ C_{10}\\
S_2(x)&=& C_{22}x^2 + C_{21}x+ C_{20}.
\label{eq:s1s2funct}
\end{eqnarray*}
The constants here are {given by}:
\begin{eqnarray*}
C_{14}&=&\sigma\left(1+\sigma\right)^2\\
C_{13}&=&\left(1+\sigma\right)^2\left(\sigma J_{11}+J_{22}\right)+2\mathrm{tr}\textbf{J}\sigma\left(1+\sigma\right)\\
C_{12}&=&\det\textbf{J}\left(1+\sigma\right)^2 + \left(\mathrm{tr}\textbf{J}\right)^2\sigma + 2\mathrm{tr}\textbf{J}\left(1+\sigma\right)\left(\sigma J_{11}+J_{22}\right)\\
C_{11}&=&2\mathrm{tr}\textbf{J}\left(1+\sigma\right)^2\det \textbf{J} + \left(\mathrm{tr}\textbf{J}\right)^2\left(\sigma J_{11}+J_{22}\right)\\
C_{10}&=&\det \textbf{J}\left(\mathrm{tr}\textbf{J}\right)^2
\label{eq:c1coeff}
\end{eqnarray*}
and
\begin{eqnarray*}
C_{22}&=&\sigma\left(1- \sigma\right)^2\\
C_{21}&=&\left(\sigma J_{11}+J_{22}\right)\left(1- \sigma\right)^2\\
C_{20}&=&J_{11}J_{22}\left(1- \sigma\right)^2.
\label{eq:c2coeff}
\end{eqnarray*}
{T}he system displays a generalized Turing instability, {determined} by the asymmetric nature of the imposed coupling if{,} after the homothetic transformation{,} the eigenvalues $D_u(\Lambda^{(\alpha)}_{Re}, \Lambda^{(\alpha)}_{Im})$
of the Laplacian operator $\boldsymbol{\mathcal{L}}$ enter the region of the complex plane $(\Lambda_{Re}, \Lambda_{Im})$ that is delimited by inequality (\ref{eq:instdom}).
 \vspace{0.5truecm}

\noindent{\textbf{{Spectral properties of circulant matrices}}}\\
A{n} $n\times n$ matrix $C$ is circulant if takes the form
\begin{equation*}
\textbf{C}=
\begin{pmatrix}
c_0 & c_{n-1} & \dots & c_{2} & c_1\\
c_1 & c_0 &  c_{n-1} & \dots & c_2\\
\vdots & c_1 &  c_0 & \ddots & \vdots\\
c_{n-2} &  &  \ddots & \ddots & c_{n-1}\\
c_{n-1} & c_{n-2} & \dots & c_1 & c_0
\end{pmatrix}.
\end{equation*}
The circulant matrix $\textbf{C}$ is fully specified by its first column, $\textbf{c}=(c_0, c_1, \dots c_{n-1})$. The other columns of $\textbf{C}$ are generated as cyclic permutations of the vector $\textbf{c}$ with offset equal to the column index. The normalized eigenvectors of a circulant matrix are given by $\phi_i=\dfrac{1}{\sqrt{n}}\left(1, \omega_i, \omega_i^2, \dots, \omega_i^{n-1}\right)$ where $\omega_i=\exp(2\pi \iota i/n)$ is the $n-$th root of the unity and the eigenvalues are $\lambda_i=c_0 + c_{n-1}\omega_i + c_{n-2}\omega_i^2 + \dots + c_1\omega_i^{n-1}$ where $i=0, 1, \dots, n-1$.\\

\end{document}


\title{{\LARGE Supplementary Material}\\\vspace*{1cm} A universal route to {pattern formation}}

\author{Malbor Asllani$^{1}$, Timoteo Carletti$^{2}$, Duccio Fanelli$^{3}$, Philip K. Maini$^{4}$ \vspace*{.25cm}}
\affiliation{$^1$MACSI, Department of Mathematics and Statistics, University of Limerick, Limerick V94 T9PX, Ireland}
\affiliation{$^2$Department of Mathematics and naXys, Namur Institute for Complex Systems, University of Namur,
rempart de la Vierge 8, B 5000 Namur, Belgium}
\affiliation{$^3$Dipartimento di Fisica e Astronomia, Universit\`{a}  di Firenze, INFN and CSDC, Via Sansone 1, 50019 Sesto Fiorentino, Firenze, Italy}
\affiliation{$^4$Mathematical Institute, University of Oxford, Woodstock Rd, OX2 6GG Oxford, UK}

%

\maketitle

\noindent{\textbf{{On the shape of the instability domain and the onset of the directed instability}}}\\

The aims of this section are twofold. {W}e will {first} provide a detailed characterization of the region  $\mathcal{D}$ of the complex plane $(\Lambda_{Re},  \Lambda_{Im})$ where the instability {eventually} takes place. {Then}, we will show that a non{-}zero inclination $\theta^*$ exists for any value of $\sigma$ strictly different from one. More specifically, we shall prove that the line tangent to the instability domain displays a slope which scales as $1/\epsilon$, for $\sigma=1 \pm \epsilon$ and {$0<\epsilon \ll 1$}. This observation yields an important consequence. It is in fact always possible to select a minimal lattice size, $\Omega$, so as to have $\theta$, the inclination of the line that connects the second eigenvalue to the origin, smaller than $\theta^*>0$. This implies{,} in turn{,} that the system can be {always} made unstable by {appropriately} setting the strength of $D_u$ (and consequently $D_v$), while keeping $\sigma$ fixed to {a} nominal value arbitrarily close to one. The instability cannot be seeded at $\sigma=1$ ($\theta^*$, being in this case zero, or, equivalently the slope of the tangent to $\mathcal{D}$, infinite), a zero measure limiting condition. 

By using a symbolic manipulator, it is possible to show that the fourth degree polynomial $S_1(x)$ admits a double root at $x_{1}=-\mathrm{tr}\textbf{J}/(D_v+D_u)$, which is a positive defined number because of the necessary assumption $\mathrm{tr}\textbf{J}<0$ (stability of the homogeneous fixed point). Factorising this latter term, we can write $S_1(x)=(x+\mathrm{tr}\textbf{J})^2Q_1(x)$ where 
\begin{equation*}
Q_1(x)=(D_v+D_u)(\mathrm{det}\textbf{J}+(D_uJ_{22}+D_vJ_{11})x+D_uD_vx^2)\, ,
\end{equation*} 
is a second order polynomial whose roots are straightforwardly determined as
\begin{eqnarray}
\label{eq:rootss1}
x_3&=&\frac{-(D_uJ_{22}+D_vJ_{11})-\sqrt{(D_uJ_{22}+D_vJ_{11})^2-4D_uD_v\mathrm{det}\textbf{J}}}{2D_uD_v}\notag \\
x_4&=&\frac{-(D_uJ_{22}+D_vJ_{11})+\sqrt{(D_uJ_{22}+D_vJ_{11})^2-4D_uD_v\mathrm{det}\textbf{J}}}{2D_uD_v}\, .\notag\\
\end{eqnarray}
If $(D_uJ_{22}+D_vJ_{11})^2-4D_uD_v\mathrm{det}\textbf{J}>0$, then $x_3<x_4$; otherwise $Q_1(x)>0$ for all $x$. In fact, $\mathrm{det}\textbf{J}>0$ owing to the stability of the homogeneous fixed point.

The characterization of $S_2(x)$ is straightforward {as it is} a second degree polynomial. Its roots can be {written}:
\begin{equation}
\label{eq:rootss2}
y_1=-\frac{J_{22}}{D_v}\quad\text{ and }y_2=-\frac{J_{11}}{D_u}\, .
\end{equation}

Without lost of generality, we can assume $J_{11}>0$ and $J_{22}<0$ (as for the case of the Brusselator model)\footnote{It is straightforward to prove that $J_{11}$ and $J_{22}$ should have different signs for the generalized instability discussed here. If $J_{11}$ and $J_{22}$ are both positive {then} the homogeneous fixed point is unstable, a setting that we ruled out by hypothesis. Conversely,  when $J_{11}$ and $J_{22}$ are both negative, there is no region in the complex plane {$\mathbb{C}$} that matches the condition set by inequality (3) in the Main Text. Hence, the two coupled species should be necessarily linked via an activator-inhibitor scheme, for the generalized instability to eventually develop (i.e. $J_{11}$ and $J_{22}$ should display opposite signs), as in the original Turing framework. Recall however that the route to instability here discussed enables one to virtually remove all constraints on the diffusion constants, as imposed under Turing's approach.}.
Hence $y_1>0$ and $y_2<0$. Observe that when $S_1$ vanishes, the domain $\mathcal{D}$ hits the $x$-axis with a vertical slope (provided $S_2$ is non{-}zero at the same time, which is generically true). On the other hand{,} if $S_2$ vanishes (but not $S_1$, which is again true in general) then the domain displays a vertical asymptote (the vertical dashed line in Fig. 2 in the Main Text).

\begin{figure*}[t]
\centering
\includegraphics[width=0.8\textwidth]{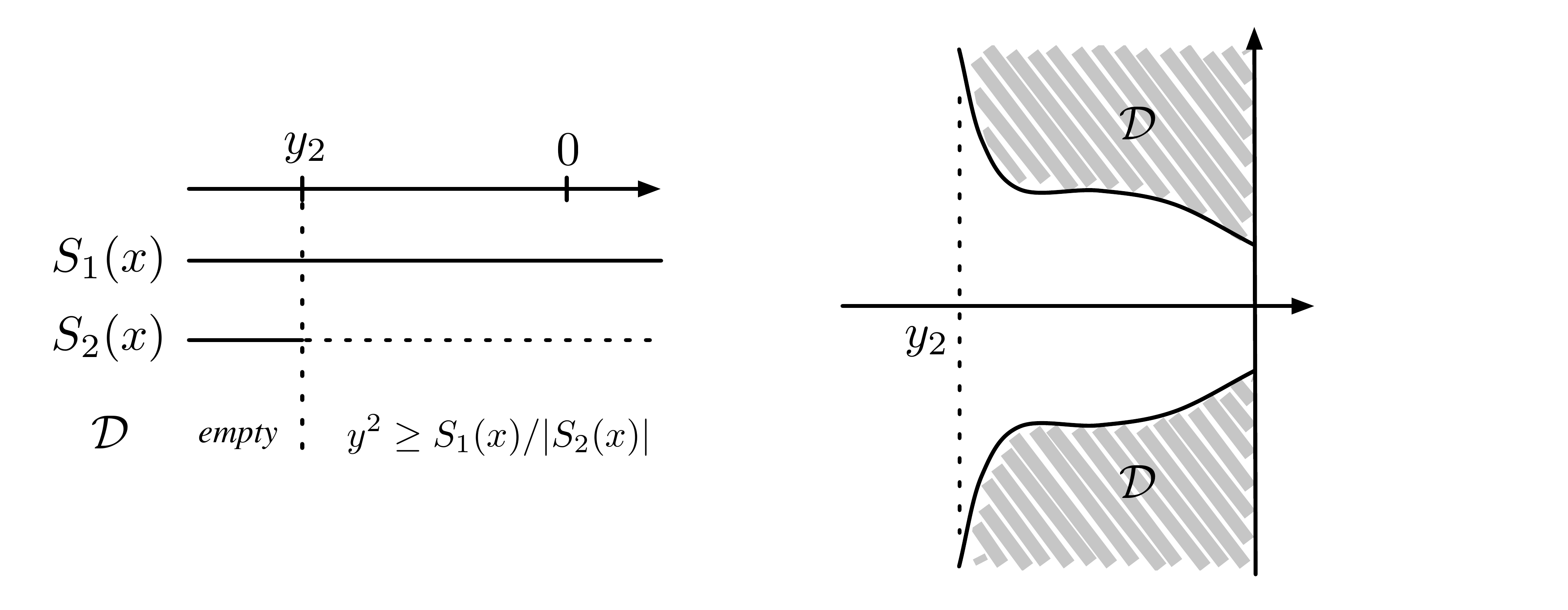}
\caption{\textbf{The domain $\mathcal{D}$: $(D_uJ_{22}+D_vJ_{11})^2-4D_uD_v\mathrm{det}\textbf{J}>0$ and $D_uJ_{22}+D_vJ_{11}<0$}. Left panel: the signs of the functions $S_1(x)$ and $S_2(x)$ for $x\leq 0$. Right panel: the corresponding domain $\mathcal{D}$ is depicted. {Note} the vertical asymptote and the fact that the domain has a limited extension on the horizontal axis. The domain never intercepts the $x$-axis. Hence, Turing patterns cannot develop on {a} symmetric support, whose Laplacian spectrum is real.}
\label{fig:case1}
\end{figure*}

\begin{figure*}[t]
\centering
\includegraphics[width=0.8\textwidth]{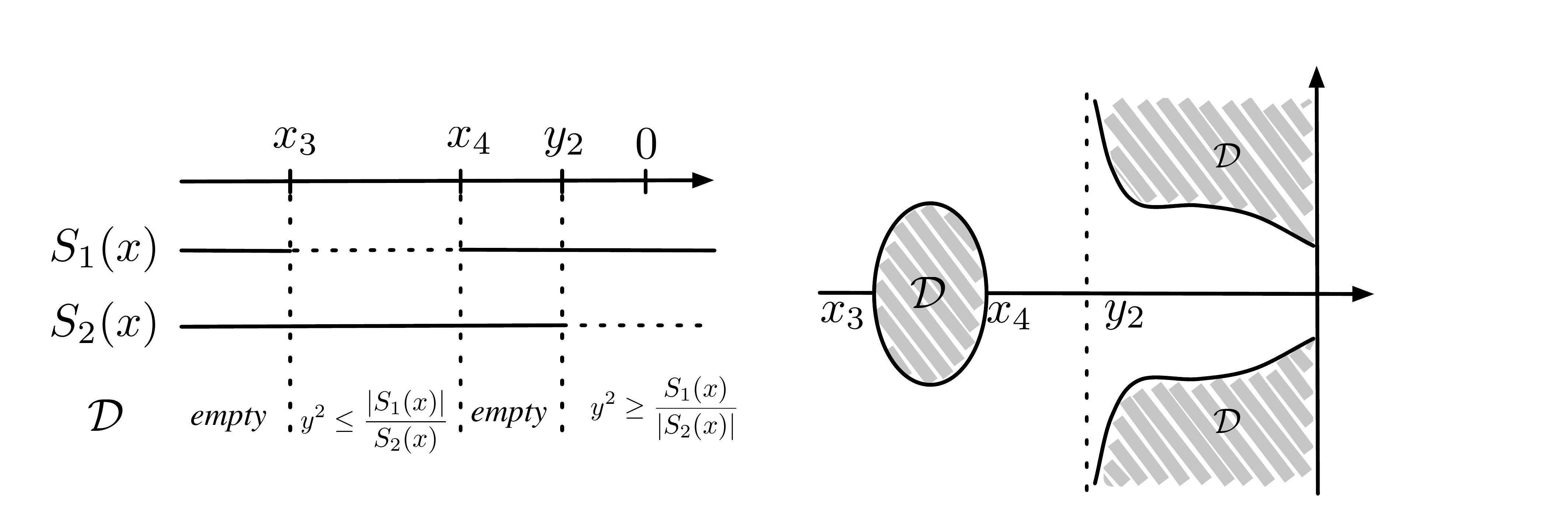}
\caption{\textbf{The domain $\mathcal{D}$: $(D_uJ_{22}+D_vJ_{11})^2-4D_uD_v\mathrm{det}\textbf{J}>0$ and $D_uJ_{22}+D_vJ_{11}>0$, $x_3<x_4<y_2$}.  Left panel: the signs of the functions $S_1(x)$ and $S_2(x)$ are characterized for $x\leq 0$. Right panel: the corresponding shape of the $\mathcal{D}$ domain is plotted. {Note} the vertical asymptote and the fact that the domain has a limited extension on the horizontal axis. The domain intercepts the $x$-axis over a finite interval. Turing patterns can hence develop on symmetric supports.}
\label{fig:case2}
\end{figure*}

\begin{figure*}[t]
\centering
\includegraphics[width=0.8\textwidth]{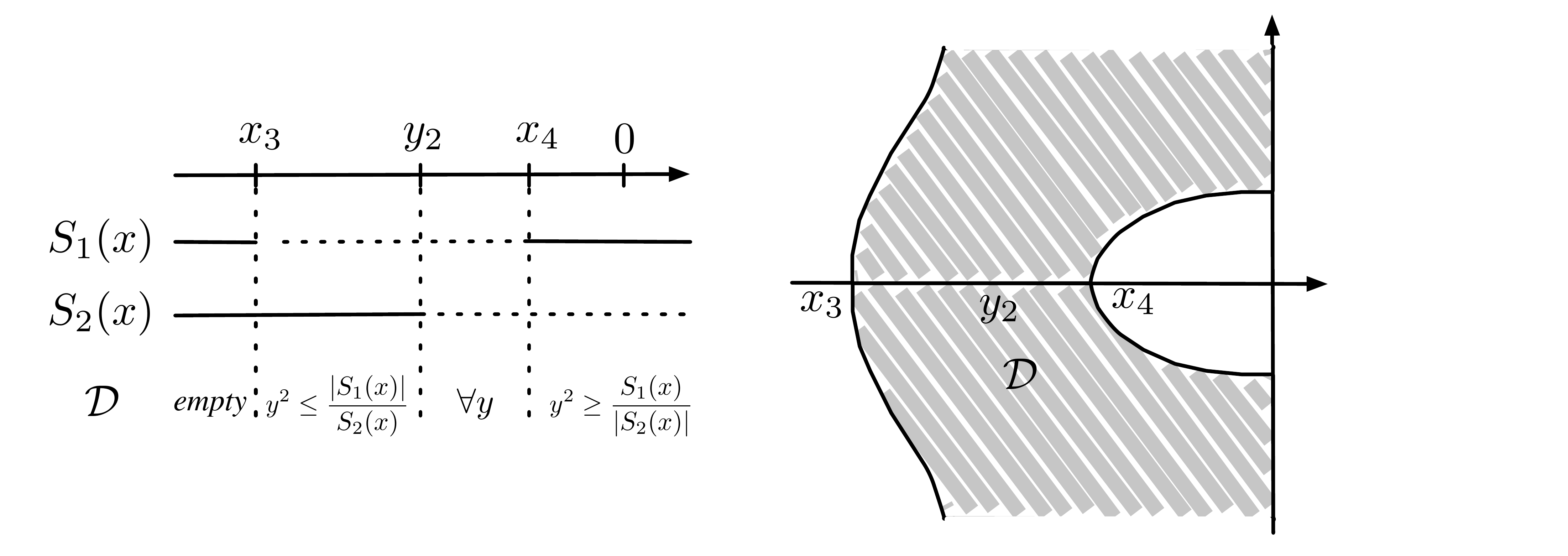}
\caption{\textbf{The domain $\mathcal{D}$: $(D_uJ_{22}+D_vJ_{11})^2-4D_uD_v\mathrm{det}\textbf{J}>0$ and $D_uJ_{22}+D_vJ_{11}>0$, $x_3<y_2<x_4$}. Left panel: the {signs} of the functions $S_1(x)$ and $S_2(x)$ for $x\leq 0$. Right panel: the corresponding shape of the $\mathcal{D}$ domain is displayed.  The domain intercepts the $x$-axis over an extended zone. Turing patterns can hence develop on symmetric supports.}
\label{fig:case3}
\end{figure*}

\begin{figure*}[t]
\centering
\includegraphics[width=0.8\textwidth]{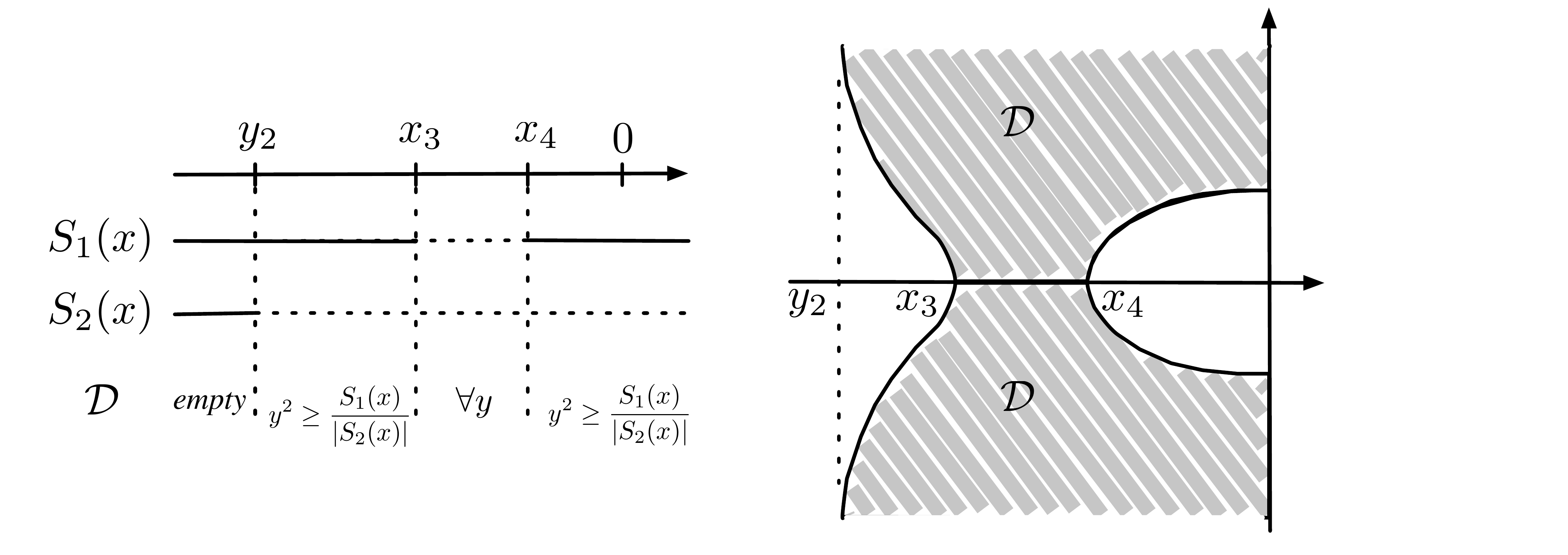}
\caption{\textbf{The domain $\mathcal{D}$: $(D_uJ_{22}+D_vJ_{11})^2-4D_uD_v\mathrm{det}\textbf{J}>0$ and $D_uJ_{22}+D_vJ_{11}>0$, $y_2<x_3<x_4$}. Left panel: the signs of the functions $S_1(x)$ and $S_2(x)$  are {shown} for $x\leq 0$. Right panel: the corresponding profile of the $\mathcal{D}$ domain is depicted. {Note} the vertical asymptote and the fact that the domain has a limited extension on the horizontal axis. The domain intercepts the $x$-axis over an extended zone, and Turing patterns are consequently possible on symmetric supports.}
\label{fig:case4}
\end{figure*}

\begin{figure*}[t]
\centering
\includegraphics[width=0.8\textwidth]{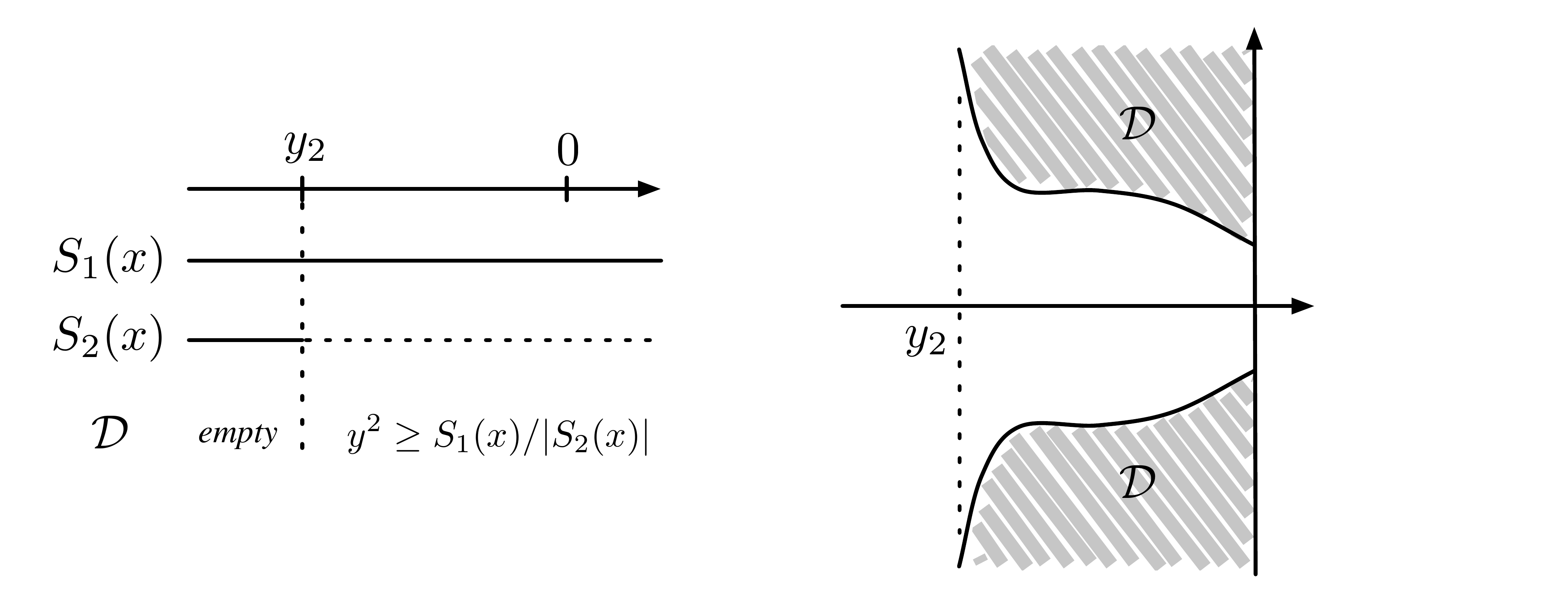}
\caption{\textbf{The domain $\mathcal{D}$: $(D_uJ_{22}+D_vJ_{11})^2-4D_uD_v\mathrm{det}\textbf{J}<0$} Left panel: the signs of the functions $S_1(x)$ and $S_2(x)$ are {shown} for $x\leq 0$. Right panel: the corresponding shape of the $\mathcal{D}$ domain is given. {Note} the vertical asymptote and the fact that the domain displays a limited horizontal extension. The domain does not intercept the $x$-axis. Turing patterns cannot set in when the system is made to evolve on a symmetric support.}
\label{fig:case5}
\end{figure*}

Collecting together the above information, we can isolate the following possible scenarios:
\begin{enumerate}
\item[\textbf{A:}] $(D_uJ_{22}+D_vJ_{11})^2-4D_uD_v\mathrm{det}\textbf{J}>0$ and $D_uJ_{22}+D_vJ_{11}<0$ (see Fig.~\ref{fig:case1}). Restricting the analysis to the relevant domain $x\leq 0$, $S_1(x)$ is always positive while $S_2(x)$ is positive for $x<y_2$ and negative otherwise. The domain $\mathcal{D}$ {therefore} displays a vertical asymptote at $x=y_2$.
\item[\textbf{B:}] $(D_uJ_{22}+D_vJ_{11})^2-4D_uD_v\mathrm{det}\textbf{J}>0$ and $D_uJ_{22}+D_vJ_{11}>0$. Here, we can identify three sub-cases, depending on the relative positions of the roots $x_3$, $x_4$ and $y_2$.
\begin{itemize} 
\item[\textbf{B.1:}] $x_3<x_4<y_2$ (see Fig.~\ref{fig:case2}). In the negative half-plane the domain $\mathcal{D}$ is composed {of} the {following} parts: two unbounded region{s} never touching the $x$-axis and a bounded domain with a finite intersection with the $x$-axis; the latter can be responsible for the emergence of Turing patterns on symmetric supports. $S_1(x)$ changes sign twice while $S_2(x)$ is positive for $x<y_2$ and negative otherwise.
\item[\textbf{B.2:}] $x_3<y_2<x_4$ (see Fig.~\ref{fig:case3}). In the negative half-plane $\mathcal{D}$ is composed {of} a single {domain} intersecting the $x$-axis. Hence Turing patterns on symmetric supports can develop. In the interval $(y_2,x_4)$ both $S_1(x)$ and $S_2(x)$ are negative, hence the inequality $S_2(x)y^2\leq -S_1(x)$ is satisfied for all $y$.
\item[\textbf{B.3:}] $y_2<x_3<x_4$ (see Fig.~\ref{fig:case4}). In the negative half-plane $\mathcal{D}$ is composed again {of} a single {domain} intersecting the $x$-axis. Hence Turing patterns on symmetric support can possibly {occur}. In the interval $(x_3,x_4)$ both $S_1(x)$ and $S_2(x)$ are negative, hence the inequality $S_2(x)y^2\leq -S_1(x)$ is satisfied for all $y$.
\end{itemize}
\item[\textbf{C:}] $(D_uJ_{22}+D_vJ_{11})^2-4D_uD_v\mathrm{det}\textbf{J}<0$ (see Fig.~\ref{fig:case5}). $S_1(x)$ is always positive while $S_2(x)$ is positive for $x<y_2$ and negative otherwise in the negative half-plane $x\leq 0$. The domain $\mathcal{D}$ displays a vertical asymptote at $x=y_2$.
\end{enumerate}

In the cases where Turing patterns cannot develop on {a} symmetric network, namely \textbf{A} and \textbf{C} (we will also consider case \textbf{B.1} because of its peculiar shape), we are interested in determining the straight line that passes through the origin and is tangent to the domain $\mathcal{D}$. Because the latter domain presents a clear symmetry $y\mapsto -y$ we will limit ourselves to considering the tangency condition for positive values of $y$. Further we shall denote {by} $y=f(x)=\sqrt{-S_1(x)/S_2(x)}$ the boundary of $\mathcal{D}$. 

\begin{figure*}[t]
\centering
\includegraphics[width=0.5\textwidth]{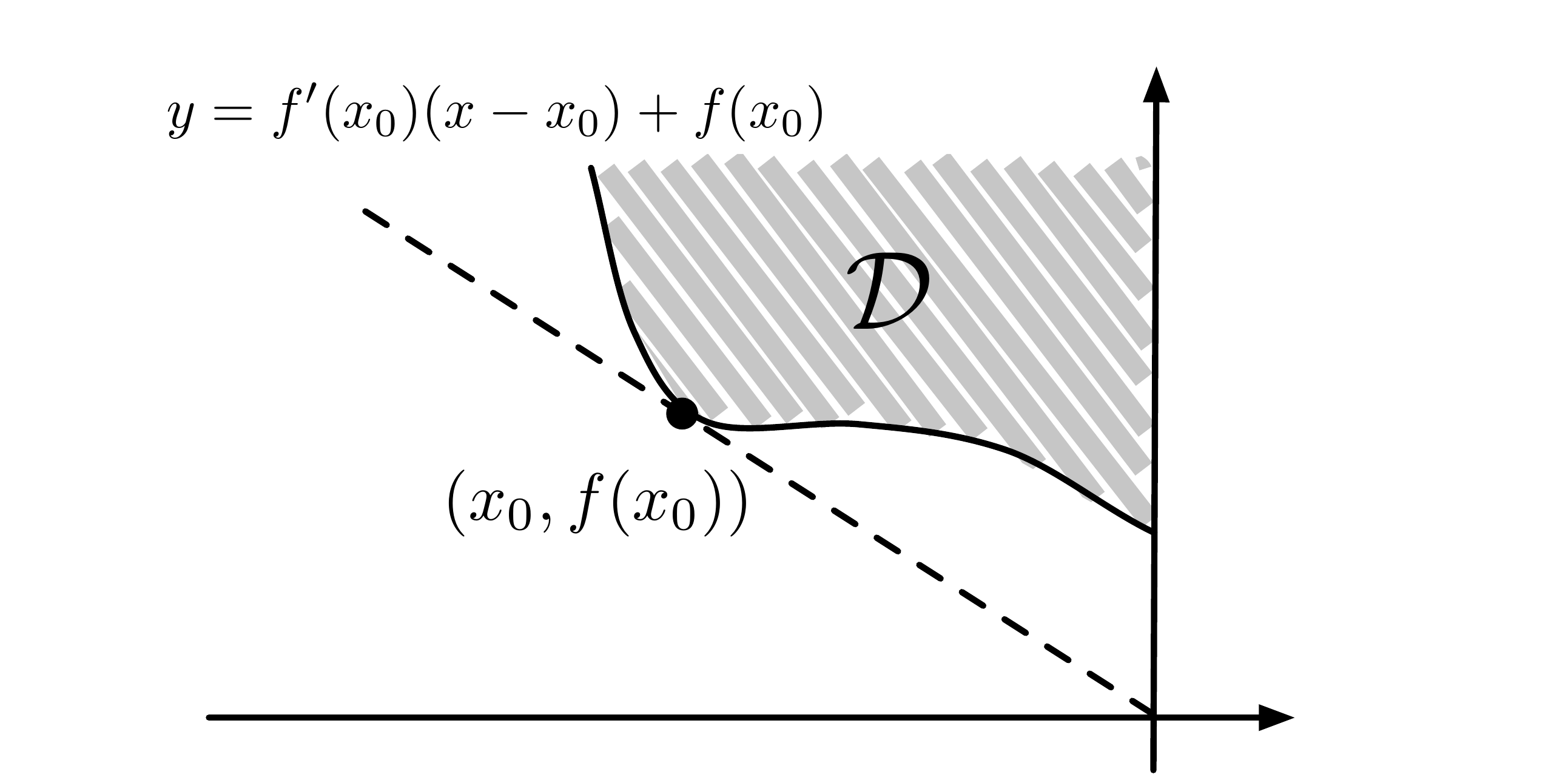}
\caption{\textbf{Tangent to $\mathcal{D}$}{: A schematic rapresentation of the tangent line} passing through the origin.}
\label{fig:tangent}
\end{figure*}

{W}e select a generic point on the boundary of $\mathcal{D}$, that is, a point with coordinates $(x_0,f(x_0))$ (see Fig.~\ref{fig:tangent}). Then, we consider the equation of the line tangent 
to the domain in $x_0$, i.e. $y=f'(x_0)(x-x_0)+f(x_0)$. Further, we let $x_0$ vary and among all the straight lines, we select the one passing through the origin, i.e. setting $x=y=0$ in the previous equation. The coordinate of such {a} point must hence satisfy: $0=-x_0 f'(x_0)+f(x_0)$.

{Using} the definition of $f(x)$, $x_0$ {must satisfy}
\begin{equation}
\label{eq:eqx0}
2S_1(x_0)S_2(x_0)-x_0(S'_1(x_0)S_2(x_0)-S_1(x_0)S_2'(x_0)=0\, ;
\end{equation}
this {appears to be} a $6$-th order polynomial. It can be shown that the coefficient of $x_0^6$ is identically equal to zero because of the definition of $C_{ij}$. We {therefore} have a $5$-th order polynomial, that we label $P_5(x_0)$. By using a symbolic manipulator, we can show that $-\mathrm{tr}\textbf{J}/(D_v+D_u)$ is a root of $P_5(x_0)$~\footnote{In fact $-\mathrm{tr}\textbf{J}/(D_v+D_u)$ is a double root of $S_1$, hence a simple root of $S_1'$. Inserting $x_0=-\mathrm{tr}\textbf{J}/(D_v+D_u)$ into Eq.~\eqref{eq:eqx0} yields the identity $0=0$.}. We are hence finally left with $P_5(x_0)=(x_0+\mathrm{tr}\textbf{J}/(D_v+D_u))P_4(x_0)$, where $P_4(x_0)$ is a polynomial of $4$-th degree, whose roots can be analytically determined by using well known formulas. 

To proceed in the analysis, we write $P_4(x_0)=e+dx_0+cx_0^2+bx_0^3+ax_0^4$. Then the four roots read
\begin{eqnarray}
\label{eq:4roots}
r_1&=&-\frac{b}{4a}-S+\frac{1}{2}\sqrt{-4S^2-2p+\frac{q}{S}}\\
r_2&=&-\frac{b}{4a}-S-\frac{1}{2}\sqrt{-4S^2-2p+\frac{q}{S}}\\
r_3&=&-\frac{b}{4a}+S+\frac{1}{2}\sqrt{-4S^2-2p-\frac{q}{S}}\\
r_4&=&-\frac{b}{4a}+S-\frac{1}{2}\sqrt{-4S^2-2p-\frac{q}{S}}\, ,
\end{eqnarray}
where
\begin{eqnarray}
\label{eq:termini}
p &=& \frac{8ac-3b^2}{8a^2}\\
q &=& \frac{b^3-4abc+8a^2d}{8a^3}\\
\Delta_0 &=& c^2-3bd+12ae\\
\Delta_1 &=& 2c^3-9bcd+27b^2e+27ad^2-72ace\\
Q &=& \left[\frac{\Delta_1+\sqrt{\Delta_1^2-4\Delta_0^3}}{2}\right]^{1/3}\\
S &=& \frac{1}{2}\sqrt{-\frac{2}{3}p+\frac{1}{3a}\left(Q+\frac{\Delta_0}{Q}\right)}\, .
\end{eqnarray}

Let us call $r_0$ any negative (real) solution. Then, we need to evaluate $f'(r_0)$, the slope of the line passing through the origin and tangent to the domain $\mathcal{D}$.
$f'(r_0)$ will depend on the set of {parameters} involved. As anticipated in the beginning of this section, we aim at characterizing the dependence of  $f'(r_0)$ on $\epsilon$, when setting 
$\sigma=1 \pm \epsilon$ and for $\epsilon \ll1$. To this end we observe that the coefficients $C_{2i}$ can be rewritten as:
\begin{eqnarray}
C_{22}&=&D_u^4(1+ \epsilon)\epsilon^2=\hat{C}_{22}\epsilon^2\\
C_{21}&=&D_u^3\left((1+ \epsilon) J_{11}+J_{22}\right)\epsilon^2=\hat{C}_{21}\epsilon^2\\
C_{20}&=&J_{11}J_{22}D_u^2\epsilon^2=\hat{C}_{20}\epsilon^2\, ,
\label{eq:c2coeff2}
\end{eqnarray}
where $\hat{C}_{2i}$ are consistently defined by the above equalities.  As a consequence, we can factor out a term $\epsilon^2$ from $S_2(x)$. This implies that $f(x)$ can be rewritten has
\begin{equation*}
f(x)=\sqrt{-\frac{S_1(x)}{S_2(x)}}=\frac{1}{\epsilon}\sqrt{-\frac{S_1(x)}{\hat{S}_2(x)}}=\frac{1}{\epsilon}\hat{f}(x)\, ,
\end{equation*}
where $\hat{S}_2(x)$ follows the definition of the coefficients $\hat{C}_{2i}$. We can hence apply the above analysis to the scaled function $\hat{f}(x)$, which is regular in the limit $\epsilon\rightarrow 0$. In particular, we can now select the negative root(s) of Eq.~\eqref{eq:eqx0} (where $S_2$ is replaced with $\hat{S}_2$ and where $\epsilon$ is set to zero), say $\hat{r}_0$. We can then compute $\hat{f}^{\prime}(\hat{r}_0)$ and conclude that ${f}^{\prime}({r}_0)=\hat{f}^{\prime}(\hat{r}_0)/\epsilon$. That is, the optimal slope gets steeper as $\epsilon$ get closer to $0$. Moreover, for any given $\epsilon$ the slope is finite, as anticipated above.\\

\bibliography{Non-normal.bib}